\documentclass{notices}
\usepackage{amsfonts,amssymb,amsmath,amscd, amsthm, xcolor}
\usepackage{graphicx}
\usepackage{url}

\theoremstyle{definition}

\title{Group-based Cryptography in the Quantum Era}

\author{
  Delaram Kahrobaei
  \affil{
    Delaram Kahrobaei is a Professor of Mathematics and Computer Science at Queens College, The City University of New York and the Honorary Chair of Cybersecurity at University of York (UK). She is a doctoral faculty at the Initiative of the Theoretical Sciences at the CUNY Graduate Center and an adjunct Professor at the New York University, Department of Computer Science and Engineering.  Her email address is dkahrobaei@gc.cuny.edu
   }
 \and
  Ram\'on Flores
  \affil{  
    Ram\'on Flores is an Associate Professor of Mathematics at the University of Seville, in Spain. His email address is ramonjflores@us.es
  }
   \and
  Marialaura Noce
  \affil{ 
  Marialaura Noce is a Research Assistant in Mathematics at University of G\"ottingen, Germany. Her email address is mnoce@unisa.it.
}
}

\begin{document}

\maketitle


\section{Introduction} 
\footnote{This paper has been accepted by the Notices of the American Mathematical Society}
Today's digital infrastructure relies on cryptography in order to ensure the  confidentiality and integrity of digital transactions. At the heart of these techniques is public key cryptography, which provides a method for two parties to communicate privately, despite the lack of any pre-arranged security keys.

These protocols mainly rely on the fact that deciphering encoded communications is tantamount to solving mathematical problems which are widely thought to be infeasible (two such examples are the factoring problem and the discrete logarithm problem). Yet we know that in the advent of large-scale quantum computers (devices that compute according to the laws of quantum mechanics), both the factoring and discrete logarithm problems are completely broken, meaning that our existing public-key cryptography infrastructure has become insecure. 

We are thus at a crossroads in terms of security: Is the security of our digital infrastructure ready for the advent of quantum computers? While security is the common goal, the mathematical theory of group theory is the common methodology. Group theory is a broad and rich theory that models the technical tools used for the design and analysis in this research.

Some of the candidates for post-quantum cryptography (PQC) have been known for years, while others are still emerging.

 Group theory, and in particular non-abelian groups, offers a rich supply of complex and varied problems for cryptography; reciprocally, the study of cryptographic algorithms built from these problems has contributed results to computational group theory. 

 \begin{center}
 \includegraphics[scale=0.25]{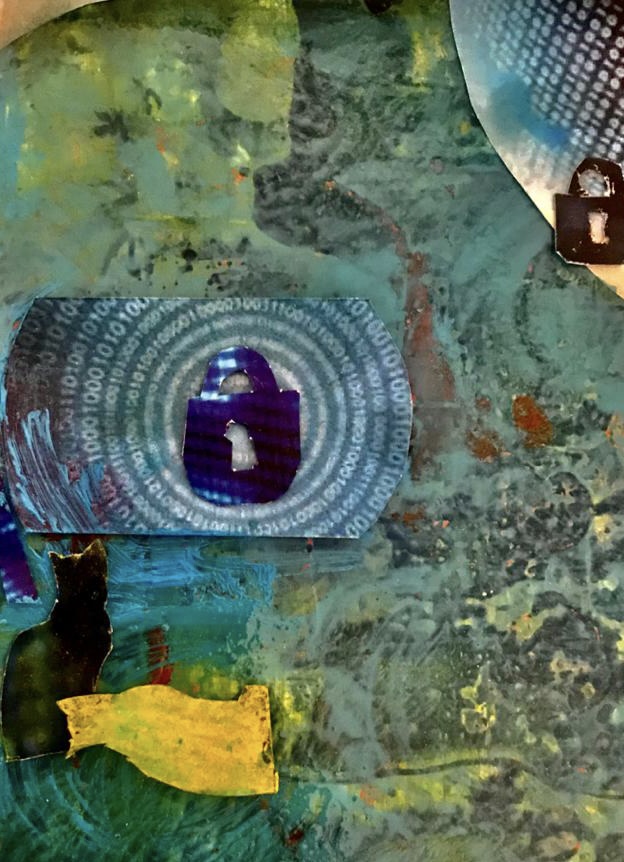}   
\end{center}

In 2015, NSA and NIST made an announcement for post-quantum cryptosystems. In July 5, 2022, the round 4 finalists were announced \cite{NIST}. Among them, the following were short-listed: lattice-based, code-based, isogeny-based and hash-based primitives.

In 2016 Anshel, Atkins, Goldfeld, and Gunnells, submitted a proposal to the NIST competition which faced several attacks by Petit et al (PKC 2017),  Blackburn et al (ASIACRYPT 2018), Ushakov et al (DCC 2019).
Recently in \cite{AaGG}, the same authors have proposed a group-based digital signature WalnutDSA$^{\text{TM}}$ which the authors claim
is safe against all those attacks and quantum-resistant.

\vspace{0.3cm}

In 1999, Anshel, Anshel, and Goldfeld \cite{AAG} proposed the commutator key-exchange protocol based on braid groups. Ko, Lee et al. proposed a non-commutative version of Diffie-Hellman using braid groups in 2000 \cite{KoLee}. 

Though braid groups were the suggested platform for both protocols, researchers have been motivated to find other suitable classes of groups for non-commutative group-based cryptography. On the other hand, in the last couple of decades, the complexity of some group-theoretic problems have been studied.

We now present a brief history of the proposed platform groups and algorithmic group theoretic problems for cryptography.

In 2004, Eick and Kahrobaei proposed polycyclic groups as a new platform for cryptography. These groups are natural generalizations of cyclic groups with more complex algorithmic theory (see Section 3.1 for more details). Grigoriev and Ponomarenko in 2004 suggested groups of matrices for a homomorphic encryption scheme.  In 2008, Ostrovsky and Skeith determined sufficient and necessary conditions for the existence of a fully homomorphic encryption scheme (FHE) over a non-zero ring if and only if there exists a FHE over a finite non-abelian simple group. Simple groups have also been proposed for hash functions by Petit and Quisquater in 2016.

In 2017, Chatterji, Kahrobaei et al studied the subgroup distortion problem in hyperbolic groups. Kahrobaei and Mallahi-Karai proposed arithmetic groups as a new platform for the same protocol in 2019. 
Since 2016 graph groups have been proposed for various cryptographic protocols by Flores, Kahrobaei, and Koberda, since several of the algorithmic problems in these groups are NP-complete which provides quantum-resistant cryptosystems (see \cite[Section 7]{FKK}). We extensively address this in section 2.2. In 2019, Kahrobaei, Tortora and Tota proposed nilpotent groups for making multi-linear maps. We conclude by mentioning that several other classes of groups were proposed in the last couple of decades for platforms for group-based cryptography. This list includes  automata groups (1991 by Garzon and Zalcstein, in 2019 by Grigorchuk and Grigoriev), Thompson group (Shpilrain and Ushakov, 2006), free metabelian groups (Shpilrain and Zapata in 2006, and Kahrobaei and Habeeb in 2012), small cancellation groups (Habeeb, Kahrobaei, Shpilrain 2012), free nilpotent $p$-groups (Kahrobaei and Shpilrain, 2016), Engel groups (Kahrobaei and Noce, 2020), and infinite pro-$p$ groups (Kahrobaei and Stanojkovski, 2021).

\vspace{0.3cm}

Next we discuss aspects that should be considered for post-quantum group-based primitive.

The security of classical cryptographic schemes such as RSA, and Diffie-Hellman are based on the difficulty of factoring large integers and of finding discrete logarithms in finite cyclic groups, respectively.
A quantum computer is able to solve the aforementioned problems attacking the security of these cryptographic algorithms. More precisely, Shor’s algorithm factors discrete logarithm problems and Grover’s algorithm can improve brute force attacks by significantly reducing search spaces for private keys. As a result, researchers are now interested in cryptography that is secure in a post-quantum world.

We recall that a subgroup $H$ of a group $G$ is \emph{hidden} by a function $f$ from $G$ to a set $S$ if it is constant over all cosets of $H$, and takes distinct values on distinct cosets.
\begin{figure}[h!]
    \centering
    \includegraphics[scale=0.55]{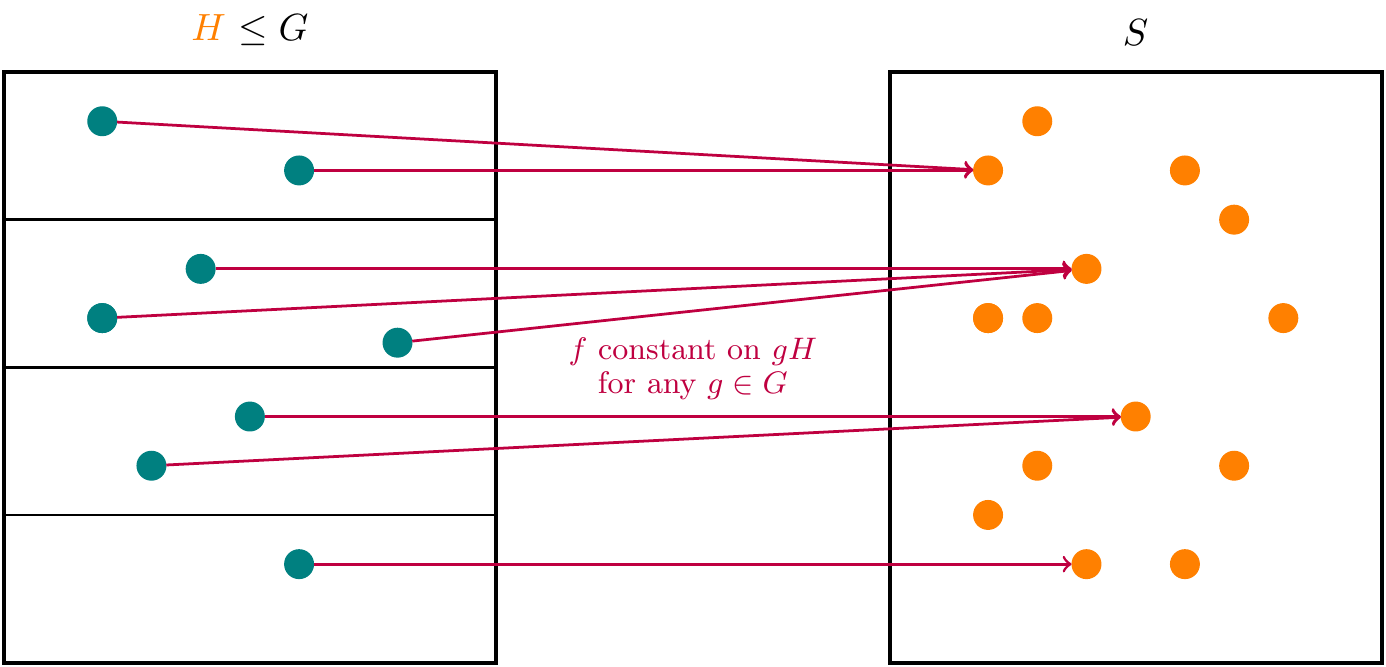}

\end{figure}

In other words, for any $g_1, g_2 \in G$, $f(g_1) = f(g_2)$ if and only if $g_1H = g_2H$.  This problem asks then whether, given a finitely generated group $G$ and an efficiently computable function $f$ from $G$ to some finite set $S$ such that $f$ is constant and distinct on left-cosets of a subgroup $H$ of finite index, we can find a finite generating set for $H$.
Given a hidden subgroup $H$, the \emph{hidden subgroup problem} (HSP for short) asks to find a generating set for $H$ using information from evaluations of $f$ via an oracle.


In summary, to analyse whether group-based cryptosystems are post-quantum one studies the relationship between the proposed algorithmic problems and the HSP,  and Grover's search algorithm for the proper parameters in the proposed platform groups. Note that cryptosystems based on NP-complete problems are not vulnerable at this time to quantum cryptanalysis.

In this paper, we present the current status and the approach of post-quantum group-based cryptography. In particular, we focus our attention to two classes of groups as platform groups for possible cryptographic protocols: polycyclic and graph groups. About the former, the complexity of algorithmic problems made polycyclic groups suitable platforms for cryptography. Likewise, graph groups are good post-quantum systems since many of the algorithmic problems presented are NP-complete. 

We remark that our treatment of the algorithmic problem in graph groups in Section \ref{Graph Groups} is more detailed, because these groups are defined more concretely and possess a very useful normal form, so their behaviour with respect to these problems is better understood than that of polycyclic groups.

We address this paper to survey several classical and novel algorithmic problems for both polycyclic and graph groups with a view towards applications to cryptography. Finally, we present a real life implementation of a combinatorial algebraic fully homomorphic encryption scheme which has been used for data analysis of encrypted medical data.
We also include a list of open problems, which we hope will guide researchers who wish to work in this  field.

\section{Platform Groups}


The study of groups mostly as combinatorial objects (using group presentation with generators and relators) is the area of group theory known as \emph{combinatorial group theory}, which has been developed in order to find solutions to the so-called \emph{decision problems} (i.e. problems with  ``yes'' or ``no'' answers).

More precisely, let $X = \{x_1, \dots, x_n\}$ and $X^{-1} = \{x_1^{-1}, \dots, x_n^{-1}\}$, where the latter is called the set of \emph{formal inverses}. The elements of $X$ and its formal inverses are called \emph{letters}, and a \emph{word} in $X\cup X^{-1}$ is a finite (possibly empty) sequence of letters of $X\cup X^{-1}$. A word $w$ in the set $X \cup X^{-1}$ is \emph{freely reduced} over $X$ if it contains no adjacent symbols $xx^{-1}$ or $x^{-1}x$. The group $G$ is a \emph{free group} with basis $X$ if $X$ is a set of generators for $G$ and no nonempty freely reduced word over $X \cup X^{-1}$ represents, as a product, the identity element of $G$ (note that the empty string represents the identity element). As an example, one can consider the group of the integers, which is the free group with a single generator.


The following three decision problems were introduced by Dehn in 1911, and are usually called the ``Dehn problems". They are defined as follows: \\
\textbf{Word Problem}: For any $g \in G$, determine if $g$ is the identity element of $G$.\\
\textbf{Conjugacy Problem}: For any $x, y \in G$, determine if $x$ and $y$ are conjugate, that is, if there exists an element $c \in G$ (a \emph {conjugator}) such that $c^{-1}xc= y$.\\
\textbf{Isomorphism Problem}: Let $G$ and $G'$ be groups given by finite presentations, determine if $G$ is isomorphic to $G'$. \\

In general decision problems are problems of the following nature: given a property $\mathcal{P}$ and an object $\mathcal{O}$, find out whether or not the object $\mathcal{O}$ has the property $\mathcal{P}$. Search problems are of the following nature: given a property $\mathcal{P}$ and an object $\mathcal{O}$ with the property $\mathcal{P}$, find something “material” establishing the property $\mathcal{P}$; for example, given two conjugate elements of a group, find a conjugator.
In other words, given a group $G$ and $a, b \in G$ where $a$ is a conjugate of $b$, the \emph{conjugacy search problem} is the problem to find an element $c \in G$ such that $c^{-1}ac= b$. The conjugation $c^{-1}ac$ is usually denoted by $a^c$.


There are many other algorithmic problems which have been used in group-based cryptography, see Section 2.2 for more examples in graph groups. 

\subsection{Polycyclic groups}
We start this section by stating the main definitions we need. A \textit{ series} of a group G is a chain of subgroups $\{1\}=G_0 \leq G_{1} \leq \dots \leq G_{n-1} \leq G_n=G$ such that each  $G_{i}$  is normal in $G_{i+1}$. A group $G$ is said to be {\em polycyclic} if it has a subnormal series $\{1\}=G_0 \leq G_{1} \leq \dots \leq G_{n-1} \leq {G_n}=G$ such that the quotient groups $G_{i+1}/G_{i}$ are cyclic. This series is called a \emph{polycyclic series}. The {\em Hirsch length} of a polycyclic group $G$ is the number of infinite factors in its polycyclic series. Though a polycyclic group can have more than one polycyclic series, it is a consequence of the Schreier Refinement Theorem that its Hirsch length is independent of the choice of series.

Every polycyclic group can be described by a polycyclic presentation of the following form:
$$
\begin {array}{lrl}
\langle g_1,\ldots,g_n \mid &
g_i^{-1}g_jg_i=u_{ij}		& \text{for} \; 1\leq i<j\leq n,\\
&
g_ig_jg_i^{-1}=v_{ij}	& \text{for} \; 1\leq i<j\leq n,\\
&g_i^{r_i} =w_{ii}	& \text{for} \; i \in I \rangle,
\end{array}
$$
where $u_{ij},v_{ij},w_{ii}$ are words in the generators $g_{i+1},\ldots ,g_n$ and $I$ is the set of indices $i \in \{1,\ldots,n\}$ such that $[G_{i+1}:G_{i}]$ is finite.\\

A group $G$ is said to be \emph{nilpotent} if and only if $G$ possess a \emph{central series}, that is, if there exists a chain of subgroups $H_0, \dots, H_n$ of $G$: 
$
\{1\}=H_0 \leq H_1 \leq \dots \leq H_n =G 
$
such that for any $i \in \{0, \dots, n\}$, $H_i$ normal in $G$ and $H_{i+1}/H_i \leq Z(G/H_i)$, where $Z(G/H_i)$ is the center of $G/H_i$.

If a group $G$ is nilpotent, the minimal length of a central series is said to be the \emph{nilpotency class} of $G$ and it is denoted by cl$(G)$.

Finally, given a prime number $p$, a group $G$ is a $p$-group if the order of every element is a power of $p$. Nilpotent groups of class 1 are abelian groups, and finite $p$-groups of order $p^a$ are nilpotent of class at most $(a-1)$.

Polycyclic groups have been always of the interest in the classical cryptography. Cyclic groups are obviously polycyclic and they have been used in the classical cryptosystems such as RSA and Diffie-Hellman.

Polycyclic groups are natural generalizations of cyclic groups with more complex algorithmic problems which provide suitable platforms for cryptography. Finitely generated nilpotent groups are polycyclic and $p$-groups are nilpotent. We discuss both applications of finite and infinite polycyclic groups here. 

Regarding the Dehn problems in polycyclic groups, the word problem can be solved efficiently, while the solution of the conjugacy problem is conjectured to be exponential time, and in particular seems not efficient. Many experiments have been run by Eick and Kahrobaei in 2004, as well as by Garber, Kahrobaei and Lam in 2013, which back up this conjecture for some classes of polycyclic groups.

There are polycyclic groups that are metabelian, as for example the group $\Sigma_3$ of permutations of three elements. Recall that a group is \emph{metabelian} if it is an extension of abelian groups. In \cite{GKM} Gryak, Kahrobaei and Martínez-P\'erez analyzed the computational complexity of an algorithm to solve the conjugacy search problem in a certain family of metabelian groups.
They proved that in general the time complexity of the conjugacy search problem for these groups is at most exponential. They also showed that for a different subfamily, namely the generalized metabelian Baumslag-Solitar groups the conjugacy search problem reduces to the discrete logarithm problem.

In \cite{GHK} Gryak, Kahrobaei and Haralick solved the conjugacy problem in certain groups via machine learning methods. These methods, improving the pre-existent machine learning and pattern recognition techniques for algorithmic problems in free groups,  allow to find heuristically the conjugate of a pair of random elements of some groups.  The groups considered are Baumslag-Solitar group $\mathcal{B}(1,2)=\langle a, b \mid bab^{-1}=a^2 \rangle$ and some non-metabelian generalisation of it, and non-virtually nilpotent polycyclic groups.

\subsubsection{Cryptographic Applications}
Polycyclic groups have many applications in group-based cryptography, see \cite{GK} for a complete survey. Such applications include the Commutators Key-Exchange Protocol based on the simultaneous conjugacy search problem and the subgroup membership search problem; the non-commutative Diffie-Hellman Key Exchange Protocol based on the conjugacy search problem; the Non-Commutative ElGamal Key-Exchange based on the power conjugacy search problem proposed by Kahrobaei and Khan;  a Key Exchange using the Subgroup Membership Search Problem; an Authentication Scheme based on the twisted conjugacy problem; authentication schemes based on semigroup actions (such as the endomorphism and the isomorphism problem) and a secret sharing scheme using the fact that there is an efficient solution for the word problem.

Below we describe a non-commutative digital signature which was proposed in 2012 by Kahrobaei and Koupparis \cite{KB} based on polycylcic groups.

\textbf{Non-Commutative Digital Signature.} Let $G$ be an infinite polycyclic group, and consider  two functions $f$ and $H$ as follows  $f\colon G\rightarrow \{0,1\}^*$, which encodes elements of the group as binary strings, and $H\colon\{0,1\}^*\rightarrow G$, known as the \emph{collision-resistant hash function}.

The functions $f$ and $H$, and the group $G$ are public and the message is signed and verified as follows:\\
\textbf{Key Generation:} The signer first chooses an element $g\in G$, whose centralizer  (the set of elements that commute with $g$) contains only the identity of $G$ and powers of $g$. The private key is an element $s \in G$ and $n \in \mathbb{N}$, where $n$ is chosen to be highly composite. The public key is $x=g^{ns}$.\\
\textbf{Signing Algorithm:} To sign a message $m$, the signer chooses a random element $t\in G$ and a random factorization  $n_in_j$ of $n$, and computes the following, where $||$ denotes concatenation:
$$
y=g^{n_it} \quad\quad h = H(m || f(y)) \quad\quad \alpha = t^{-1}shy.
$$
The signature $\sigma=\langle y, \alpha,n_j\rangle$ and the message $m$ are then sent to the message recipient.\\
\textbf{Verification:} To verify, the recipient computes $h' = H(m || f(y))$, and accepts the message as authentic if and only if the following equality holds:
$y^{n_j\alpha} = x^{h'y}.$

The security of the signature scheme is based on the collision resistance of the hash function and the hardness of the conjugacy search problem in $G$ in the platform group.


\textbf{Multilinear maps.}
In the last decades, multilinear maps have attracted attention in cryptography. In 2003, Boneh and Silverberg  proposed  multilinear maps in cryptography, exploring in particular how to build these maps. In 2017 Mahalanobis and Shinde presented a novel  bilinear cryptosystem in  groups of nilpotency class 2. In order to explore more deeply these maps, Kahrobaei, Tortora and Tota proposed multilinear cryptosystem using identities in nilpotent groups in 2019. Recently, Kahrobaei and Stanojkovski proposed pro-$p$ groups in general form for such maps and analyzed the security \cite{KS}. 

In order to explain the aforementioned results, we give a couple of useful definitions.  We first recall that  given a group $G$ and $x_1, \dots, x_n \in G$ a simple commutator of weight $n>1$ is defined recursively by the rules $[x_1,x_2]=x_1^{-1}x_2^{-1}x_1x_2$, and
$$
[x_1, x_2, \dots, x_n] = [[x_1, \dots, x_{n-1}], x_n]
$$
if $n>2$. Sometimes we will use the following shorthand notation
$$
[x,_n y]=[x, y, \overset{n}{\dots}, y].
$$
Let now $n$ be a positive integer and $G$ an arbitrary group. A map $e: G^n  \to G$ is said to be a \emph{multilinear map} if for any $g_1, \dots, g_n, \in G$ and any $a_1, \dots, a_n \in \mathbb{Z}$ we have 
    $$
    e(g_1^{a_1}, \dots, g_n^{a_n}) = e(g_1, \dots, g_n)^{a_1 \cdots a_n}.
    $$
Moreover, we say that the map $e$ is non-degenerate if there exists $g\in G$ such that $e(g, \dots, g) \neq 1$.

If furthermore $G$ is a nilpotent group, there are additional properties for multilinear maps. So let $G$ be a nilpotent group of class $n > 1$ and  $g_1, \dots, g_n$ elements of $G$. One can easily prove by induction on $n$ that for any $a_1, \dots, a_n \in \mathbb{Z}$ the following identity holds:
\begin{equation}\label{identitynilpotent}
 [g_1^{a_1}, \dots, g_n^{a_n}] = [g_1, \dots, g_n]^{\prod_{i=1}^n a_i}.
\end{equation}
Hence if $G$ is nilpotent, the map $e$ 
\begin{align*}
 e: G^{n} & \to G \\
 (g_1, \dots, g_{n}) & \mapsto [g_1, \dots, g_{n}]
\end{align*}
is a multilinear map. In addition, if we fix $x \in G$, we can construct another multilinear map $f$ given by
\begin{align*}
 f: G^{(n-1)} & \to G \\
 (g_1, \dots, g_{n-1}) & \mapsto [x, g_1, \dots, g_{n-1}].
\end{align*}
If the multilinear map is non-degenerate, then one can propose multilinear cryptosystems using identities in nilpotent groups in multiparty key-exchange protocols, in which the security is based on the discrete logarithm problem. The protocol presented by Kahrobaei, Tortora and Tota reads as follows.

Let $n$ be a positive integer, and suppose that the public group $G$ is nilpotent  of class $n + 1$, but not $n$-Engel. We recall that a group $G$ is an  \textit{$n$-Engel} group if there exists $n \geq 1$ such that $[x, {}_ny] = 1$, for all $x, y \in G$. Consider then $n + 1$ users $A_1, \dots A_{n+1}$ that wish to agree on a shared secret key. Each user $A_j$ selects a private integer $a_j \neq 0$, computes $g^{a_j}$, and sends it to the other users. Then we are in the following situation:
\begin{itemize}
    \item The user $A_1$ computes $[x^{a_1},g^{a_2}, \dots g^{a_{n+1}}]$.
    \item For $j=2, \dots, n$, the user $A_j$ computes $[x^{a_j}, g^{a_1}, \dots, g^{a_{j-1}}, g^{a_{j+1}}, \dots, g^{a_{n+1}}]$. 
    \item The user $A_{n+1}$ computes $[x^{a_{n+1}},g^{a_1}, \dots g^{a_{n}}]$.
\end{itemize}

Since  the identity \eqref{identitynilpotent}  holds in all nilpotent groups, all elements computed by the users are equal to 
$$
k=[x,_n g]^{\prod_{j=1}^{n+1} a_j},
$$
where $k$ is the shared key.

In \cite{KS}, Kahrobaei and Stanojkovski propose a new protocol employing multilinear maps for an arbitrary number of users. This protocol is a \emph{Non-Interactive Key Exchange} in which the users  agree on a symmetric shared key without any interaction, and for this reason is said to be \emph{non-interactive}. Note that one of the most known Non-Interactive Key Exchange schemes (NIKE, in what follows, for short) is the Diffie and Hellman key-exchange protocol over cyclic groups. 

Let $n$ be an integer greater than 2, and let $G$ be a nilpotent group of nilpotency class $n$. We set:

\begin{itemize}
    \item \textbf{Public:} $g_1, \dots, g_n \in G$
\item \textbf{Users:} $A_1, \dots A_{n+1}$, who choose an integer $a_i \in \mathbb{Z}$.
\item \textbf{Private keys:}  $a_1, \dots a_{n+1}$
\item \textbf{Public shared data:} $g_i^{a_j}$ with $1 \leq i \leq n$ and $1 \leq j \leq n+1$
\item \textbf{Shared secret key:} $[g_1^{a_{n+1}}, g_2^{a_2}, \dots, g_n^{a_n}]^{a_1} = [g_1, \dots, g_n]^{\prod_{i=1}^{n+1} a_i}$, which can be computed from the shared data since the commutator is a multilinear map. 
\end{itemize}

The security of the above protocol is based on the difficulty to recover the shared key. For a finite $p$-group this can be reduced to solve the Discrete Logarithm Problem in a cyclic group of order $p^a$, which is known to be classically hard. We observe that the case $n = 2$ already was analyzed by Mahalanobis and Schinde in 2017.

Motivated by the use of the protocol above in a more general context and for an arbitrary number of users, Kahrobaei and Stanojkovski in 2021 employed infinite pro-$p$ groups. More precisely, consider a non-nilpotent profinite $p$-group $\overline{G}$ with $n \geq 2$ an integer. It is known that then $G$ has a finite quotient of nilpotency class $n$ and so, over $G$, one can construct a key exchange protocol between $n + 1$ users. Kahrobaei and Stanojkovski proved that such a group $\overline{G}$ exists and it can be used as a platform for an arbitrary number of users. 

It is worth mention here the definition of the \textbf{Generalised Discrete Logarithm Problem}, as it is connected to the security of the above mentioned multilinear maps.
Let $G$ be a finite group. Given $x, y \in G$, the Discrete Logarithm Problem (in the remainder DLP for short) is the problem to find whether there exists a positive integer $a$ such that $x^a=y$. Notice that this is usually defined in the setting of cyclic groups because the Discrete Logarithm exists for all elements and all nontrivial bases. The DLP can be generalized to several components as follows. Let $\textbf{x} = (x_1, \dots, x_n)$ be a tuple of elements such that $G = \langle x_1, \dots, x_n \rangle$. Given $y \in G$, the \emph{Generalised Discrete Logarithm Problem} of $y$ with respect to $\textbf{x}$ is to find $a_i$ such that $y$ can be written uniquely as $$
\textbf{x}^{\textbf{a}} = x_1^{a_1} \dots x_n^{a_n}=y,
$$
where $0 < a_i < |x_i|$ for any $i$.

In 2011 Sutherland presented a generic algorithm to compute Generalised Discrete Logarithms in every finite abelian $p$-group $G$ by using some direct methods to compute a basis for $G$ \cite{Sutherland}. It is an interesting problem to find the complexity of this problem for any finite $p$-group.

\textbf{Semidirect Product Key-exchange Protocol.} Habeeb, Kahrobaei, Koupparis, Shpilrain in 2013 proposed a key-exchange protocol using semidirect product \cite{ACNS}. A few platforms have been proposed, for example, in \cite{CiE}, free nilpotent $p$-groups were proposed. Recently, Battarbee, Kahrobaei, Perret, and Shahandashti gave the first dedicated security analysis of Semidirect Discrete Logarithm Problem. In particular, they provide a connection between Semidirect Discrete Logarithm Problem and group actions, a context in which quantum subexponential algorithms are known to apply \cite{battarbee2022subexponential}.\\

Here we give general ideas of this protocol.
Let $G$ be a (semi)group. An element $g\in G$ is chosen and
made public as well as an arbitrary automorphism $\phi\in Aut(G)$
(or an arbitrary endomorphism $\phi\in End(G)$). Bob chooses a
private $n\in \mathbb{N}$, while Alice chooses a private $m\in
\mathbb{N}$. 
Both Alice and Bob are going to work with elements of the form $(g, \phi^r)$, where $g\in G, r \in \mathbb{N}$. Note that two elements of this form are multiplied as follows:  $(g, \phi^r) \cdot (h, \phi^s) = (\phi^s(g) \cdot h, \phi^{r+s})$.

\begin{itemize}
\item Alice computes $(g, \phi)^m = (\phi^{m-1}(g) \cdots \phi^{2}(g) \cdot \phi(g) \cdot g, \phi^m)$ and sends {\bf only the first component} of this pair to Bob. Thus, she sends to Bob {\bf only} the element $a = \phi^{m-1}(g) \cdots \phi^{2}(g) \cdot \phi(g) \cdot g$ of the (semi)group $G$.
\item Bob computes $(g, \phi)^n = (\phi^{n-1}(g) \cdots \phi^{2}(g) \cdot \phi(g) \cdot g, \phi^n)$ and sends {\bf only the first component} of this pair to Alice. Thus, he sends to Alice {\bf only} the element $b = \phi^{n-1}(g) \cdots \phi^{2}(g) \cdot \phi(g) \cdot g$ of the (semi)group $G$.
\item Alice computes $(b, x) \cdot (a, \phi^m) = (\phi^m(b) \cdot a, x  \cdot \phi^{m})$. Her key is now $K_A = \phi^m(b) \cdot a$. Note that she does not actually ``compute" $x \cdot \phi^{m}$ because she does not know the automorphism $x=\phi^{n}$, and also recall that it was not transmitted to her, but she does not need it to compute $K_A$.
\item Bob computes $(a, y) \cdot (b, \phi^n) = (\phi^n(a) \cdot b, y \cdot \phi^{n})$. His key is now $K_B = \phi^n(a) \cdot b$. Again, Bob does not actually ``compute" $y \cdot \phi^{n}$ because he does not know the automorphism $y=\phi^{m}$.
\item Since $(b, x) \cdot (a, \phi^m) = (a, y) \cdot (b, \phi^n) = (g, \phi)^{m+n}$, we should have $K_A = K_B = K$, the shared secret key.
\end{itemize}

The proposed algorithmic problem on which the security of this scheme is based is a cousin of the Computational Diffie-Hellman problem. There is no known reduction from this problem to the DLP.

\subsection{Graph Groups}
\label{Graph Groups}

Graph groups (also called  partially commutative groups, semifree groups, right-angled Artin groups, or simply RAAGs in the literature), were defined by Baudisch (1977), as a kind of interpolation between free groups and and free abelian groups. They admit a presentation where the only relations are commutativity relations which are codified in a finite simplicial graph, see the definition below. The fact that these groups are defined by means of a graph implies that there is a tight connection between algorithmic graph theoretic problems and group theoretic problems for graph groups. Since the graph theoretic problems have been of central importance in Complexity Theory, it is natural to consider some of these graph theoretic problems via their equivalent formulation as group theoretic problems about graph groups.

In general, given a property of a graph, it is easy to figure out the corresponding group-theoretic property of the associated graph group, via the graph that defines it. However, given an intrinsic property of the graph group (i.e. not depending on any particular set of generators), it is usually hard to characterize the corresponding graph property, and not always possible. For example, if a graph is not connected it is nearly immediate that the associated graph group decomposes as a free product, but the reciprocal result is a highly non-trivial theorem. Since the eighties, an important line of research has been developed in order to model group-theoretic properties of graph groups in the terms of properties of the graph. 



The previous approach permits in particular to convert graph theoretic problems for finite graphs into group theoretic ones for graph groups. Motivated by the fact that some of these group theoretic problems can be used for cryptographic purposes, such as authentication schemes, secret sharing schemes, zero-knowledge proofs, hash functions and key exchange protocols, Flores, Kahrobaei and Koberda have considered these groups as a promising platform for several cryptographic schemes (see \cite{FKK}, \cite{FKK-C}, \cite{FKK-H}, \cite{FKK-E}). It is important, in this sense, that the good knowledge of the group-theoretic structure of these groups (normal forms, centralizers, automorphisms, subgroups, etc.) make their algorithmic properties very tractable.

Next we will define rigorously graph groups and describe some of their main features from the cryptographic point of view.

\subsubsection{Main definitions}
Here we define graph groups: Let $\Gamma$ be a finite simplicial graph. We write $V = V (\Gamma)$ for the finite set of vertices and $E(\Gamma) \subset V \times V$ for the set of edges, viewed as unordered pairs of vertices.  The \emph{graph group} on $\Gamma$ is the group
$$A(\Gamma) = \langle V|[v_i, v_j] =1 \text{ whenever } (v_i, v_j) \in E \rangle.$$
In other words, $A(\Gamma)$ is generated by the vertices of $\Gamma$, and the only relations are given by commutation of adjacent vertices. For example, if $\Gamma$ is just an edge, then $A(\Gamma)$ is $\mathbb{Z}\times\mathbb{Z}$, the free abelian group in two generators.

The previous presentation is frequently called a \emph{standard presentation} of the graph group, and the generators the \emph{standard generators} or \emph{Artin generators}. The number of vertices of the graph is the \emph{rank} of the group. It is clear by the definition that the graph determines the group, and by the work of Droms (1987), the converse is also true. In the following, given a graph $\Gamma$, we will denote by $A(\Gamma)$ is the associated graph group, and conversely, given a graph group $A$, we will denote by $\Gamma (A)$ its associated graph. We will always assume that the graphs that appear in this section are finite.

\subsubsection{Algorithmic problems}
Next we will comment on the main algorithmic problems in the context of graph groups, and the different solutions that have been given to them throughout the years.

\textbf{Word problem}. Servatius, using normal forms, gave in 1987 a first solution of the word problem for graph groups, although he paid no attention to the complexity of the construction of the normal forms. A bit later, Wrathall (1988) used good properties of a presentation of the monoid of positive words to prove that the word problem is solvable for graph groups in linear time. Expanding these methods, Liu-Wrathall-Zeger (1990) established that the generalised word problem (i.e. given two words $x$ and $y$ in the group, check if some power of $x$ is equal to some power of $y$) is also solvable in linear time, where the argument of linearity is the length of the
product.

\textbf{Conjugacy problem}. The approaches just described by Servatius and Liu, Wrathall, and Zeger were also useful to prove respectively that the conjugacy problem for graph groups is solvable in linear time. More recently (2009), Crisp, Godelle, and Wiest use a version of the Viennot and pyramidal pilings for graph groups to construct a new normal form. In this way, they are able to prove that in fact the complexity of the conjugacy problem in this context is linear in terms of the sum of the lengths of the elements involved. The technique consists in constructing the normal forms out of the corresponding pilings, and comparing them. In this way these authors also prove the linearity of the conjugacy problem for an important family of subgroups of graph groups, namely fundamental groups of Haglund-Wise special (or virtually special) cube complexes.
It is worth mentioning here that the richness of the subgroup structure of graph groups gives rise to finitely presented examples. In general, the corresponding problem is not solvable for subgroups of graph groups, not even if they are finitely generated. Bridson (2013) found a finite index subgroup $P$ of a graph group $A$ and a finitely generated free group $F$ such that the Isomorphism problem is not solvable for the finitely presented subgroups of $P\times P\times F$, and then for subgroups of $A\times A\times F$, which is a graph group. 

\textbf{Subgroup Isomorphism problem}. Recall that the Subgroup Isomorphism problem asks if given two groups $G$ and $H$ by presentations, can $G$ be embedded as a subgroup of $H$ or not. If $G$ and $H$ are graph groups given by standard presentations, a sufficient condition for $G$ to be a subgroup of $H$ is that the graph associated to $G$ is an induced subgraph of the graph associated to $H$ (i.e. a subgraph such that if $v$ and $w$ are vertices of $G$ and the edge $vw$ belongs to $H$, then it also belongs to $G$). It is known that this problem is NP-complete in general. However, in principle it would be possible to always find an embedding $G<H$ that does not involve the graph, as for example any embedding $F_3<F_2$ of free groups. But this is not possible: using the techniques of the previous paragraph, Bridson also proved that there is no general solution for the Subgroup Isomorphism problem in graph groups.

\textbf{Group Homomorphism problem}. The general version of the Group Homomorphism problem asks if given two groups $G$ and $H$, is there a nontrivial homomorphism $G\rightarrow H$. For example, if $G$ is simple and $H$ does not contain a copy of $G$, the answer is clearly negative. In turn, recall that given two graphs $\Gamma_1$ and $\Gamma_2$, a homomorphism $f:\Gamma_1\rightarrow \Gamma_2$ is an assignation that takes vertices to vertices and edges to edges. It is easy to see that not every homomorphism between graph groups can be realized as a homomorphism between the associated graphs, even if it takes standard generators to standard generators. For example, the first projection $\mathbb{Z}^2\rightarrow \mathbb{Z}$ should be given by a homomorphism $K_2\rightarrow K_1$, which does not exist. Here $K_n$ denotes the complete graph in $n$ vertices, also called $n$-clique.

Hence, from the point of view of cryptography, it is very useful to consider only the homomorphisms between two graph groups $G_1$ and $G_2$ with standard presentations that take standard generators of the first to standard generators of the second, and such that two commuting standard generators are taken to two different standard generators that commute. Now if we are restricted to this case, the problem of finding such a homomorphism between $G_1$ and $G_2$ is equivalent to the Graph Homomorphism problem for the associated graph, which is known to be an $NP$-complete coloring problem (Johnson, 1979).

\textbf{The Membership problem}. Given a group $H$ and a subgroup $K<H$ and presentations of $H$ and $K$, the Membership problem consists in deciding if an element of $H$ belongs to $K$. Recall that given a presentation of a group $G$, the norm $|g|$ of an element of $G$ is defined as the minimal length of a word (in the given generators and their inverses) that represents $g$. Then, given two elements $g$ and $h$ in the group, the distance between $g$ and $h$ is defined as the norm of $g^{-1}h$. In this way a metric on $G$ is defined, called the \emph{word metric}. For example, in the free group $F_2=\langle a,b\rangle$, the distance between $ab$ and $ab^{-1}a^2$ is $|b^{-1}a^{-1}ab^{-1}a^2|=|b^{-2}a^2|=4$. Now consider presentations of groups $K$ and $H$, the associated word metrics $d_K$ and $d_H$ associated to the presentations and a monomorphism $i:K\hookrightarrow H$. Then $K$ is \emph{undistorted} in $H$ if the embedding is a quasi-isometry, i.e. there exist constants $A\geq 1$, $B\geq 0$ such that for every $x,y\in K$ we have $$\frac{1}{A}d_K(x,y)-B\leq d_H(i(x),i(y))\leq Ad_{K}(x,y)+B.$$ Otherwise $K$ is said to be \emph{distorted} in $H$. For every $h\in K$, we can define the \emph{distortion function} $D:\mathbb{N}\rightarrow \mathbb{N}$ as $D(n)=\textrm{max}\{|h| \textrm{ such that} |i(h)|\leq n\}.$

It was proved by Flores, Kahrobaei and Koberda \cite{FKK} that if $G$ is a group where the word problem is solvable in at most exponential time, the Membership problem is so for every finitely generated undistorted subgroup. In particular, we have seen above that the word problem is in fact linear for graph groups, so they fit in this framework. Moreover, in that paper it is shown that there exists a graph group $G$ and a subgroup $H<G$ isomorphic to the fundamental group of a compact surface such that a) its distortion function has exponential growth, and b) its Membership problem is also solvable in at most exponential time (it could be even polynomial). On the contrary, Bridson has described examples of distorted subgroups of graph groups for which the Membership problem remains unsolvable.

\textbf{The Geodesic problem}. For a given presentation of a group $G$, a word in the generators is said to be $\emph{geodesic}$ if its number of letters coincides with the length of the element of the group that it represents; in other words, it is a shortest word in these generators representing the element. There are several classical algorithmic problems involving geodesics and length. The Geodesic problem is, given an element $g$ in a group $G$, to find a geodesic word that represents the elements. The Geodesic length problem consists in computing the length of an element in the group. There is a bounded version of the latter, where it is intended to determine for a natural $k$ if the length of an element is smaller or equal to $k$. It is known that these three problems have the same complexity (as each of them is reducible to each other), and in the case of graph groups this complexity is polynomial by a result of Holt and Rees (2013).


\textbf{The Decomposition problem}. Recall that given two graphs $\Gamma_1=(V_1,E_1)$ and $\Gamma_2=(V_2,E_2)$, their join is the graph whose vertex set is $V_1\cup V_2$ and whose edges set is given by $E_1\cup E_2$ and all the possible edges that start in $V_1$ and end at $V_2$. Then it is known (Servatius, 1989) that a graph $\Gamma$ can be decomposed as a nontrivial join if and only if the graph group $A(\Gamma)$ decomposes as a nontrivial direct product. In \cite{FKK} is described an algorithm (probably known previously) which decomposes any graph as a join of graphs which in turn cannot be further decomposed. This algorithm stops in polynomial time, and this proves that decomposing a group as a direct product of indecomposable subgroups can be also solved in polynomial time, provided we have an standard presentation of the group.



\textbf{Hamiltonicity}. Flores, Kahrobaei and Koberda defined in \cite{FKK-H} the concept of Hamiltonian vector space. Consider a triple $(V,W,q)$ where $V$ and $W$ are vector spaces over a field $F$, $q:V\times V\rightarrow W$ an (anti-)symmetric bilinear
pairing on $V$. It is said that $(V, W, q)$ is a \emph{Hamiltonian vector space} if whenever $(w_1,\ldots,w_n)$ is a
basis for $V$ then there is a permutation $\sigma$ of $n$ elements such that for all $1\leq i<n$, we have $q(w_{\sigma(i)},w_{\sigma(i+1)})\neq 0$, $q(w_{\sigma(n)},w_{\sigma(1)})\neq 0$. Given a graph group $A$, the Hamiltonicity of the triple $(H^1(A,F),H^2(A,F),\cup)$, where $H^n(A,F)$ denotes the $n$-th homology group of $A$ with coefficents in $F$ and $\cup$ denotes the cup product, is an invariant of the isomorphism type of the group. Then it is proved in the aforementioned paper that the fact that this vector space is Hamiltonian is equivalent to the Hamiltonicity (in the classical sense) of $\Gamma (A)$. Then, given a graph group $A$, the problem of determining if $(H^1(A,F),H^2(A,F),\cup)$ is Hamiltonian is NP-complete. Observe that the definition of Hamiltonian vector space models algebraically the property of possessing a Hamiltonian cycle; an analogous result is valid, \emph{mutatis mutandis}, when considering Hamiltonian paths instead of cycles.

\subsubsection{Cryptographic applications}

In this section we review several cryptographic applications of graph groups and protocols that have been developed out of them.

\textbf{Secret sharing schemes}. Basing on previous work by Habeeb, Kahrobaei and Shpilrain and Shamir, Flores and Kahrobaei proposed in 2016 secret sharing schemes using graph groups, which rely on the fact that the word problem in these groups is solvable in linear time. To illustrate the ideas that are used, we describe one scheme of each type. We start with a sharing scheme, which uses decisively that the word problem can be solved in linear time in graph groups. The idea of the scheme is that the dealer distributes a $k$-vector $C=(c_1,c_2,\ldots, c_k)$ of 0's and 1's among $n$ participants, making sure that the vector can only be totally reconstructed if all participants share their information.

So let us describe the scheme. First, a set $\{x_1,\ldots,x_m\}$ of public generators is selected.

\begin{itemize}

\item Each participant receives secretly from the dealer a set of commutators $R_j$ of the generators in $X$ (and their inverses), so the participant $P_j$ possesses the graph group $G_j=\langle X\textrm{ } | \textrm{ } R_j\rangle$

\item The vector $C$ is written by the dealer as a $mod$ 2 sum $C=\sum_{i=1}^n C_i$ of $n$ $k$-vectors. We denote by $c_{ij}$ the $i$-th entry of $C_j$. The vector $C_j$ will be the secret of the participant $P_j$.

\item In turn, the participant $P_j$ also receives publicly a set of words $\{w_{1j},\ldots,w_{kj}\}$, selected in such a way that the element represented by $w_{ij}=1$ in $G_j$ if $c_{ij}=1$ and $w_{ij}\neq 1$ otherwise.

\item Using that the word problem in graph groups can be solved efficiently, each participant $P_j$ checks the triviality or not of the words $\{w_{1j},\ldots,w_{kj}\}$, and in this way he gets the vector $C_j$.

\item Finally, the sum of the vectors $C_j$ reconstructs the original message.

\end{itemize}

Another protocol developed in \cite{FKK} uses the Decomposition problem. For each of $n$ participants $\{P_1,\ldots,P_n\}$, the dealer distributes through a secure channel a right-angled Artin group $A(\Gamma_i)$. As the Decomposition problem is efficiently solvable, the participant $P_i$ can compute a bit $b_i$ such that $b_i=0$ decomposes as a non-trivial join, and $b_i=1$ otherwise. Let now $f(x)$ be the only monic polynomial of degree $n$ such that $f(i)=m_i$. Then the polynomial can be reconstructed out of these values, and the secret key is $f(0)$.

\textbf{Authentication schemes}. 
Flores and Kahrobaei in 2016 proposed authentication schemes using graph groups as platforms. The authentication protocols depend on the complexity of the Group Homomorphism problem (which is NP-complete) and the Subgroup Isomorphism problem (which is NP-complete for certain classes of graph groups).

Let us now describe the authentication protocol.

\begin{itemize}

\item Alice's public key is given by two graph groups $G_1=\langle S_1 \textrm{ } | \textrm{ } R_1 \rangle$ and $G_2=\langle S_2 \textrm{ } | \textrm{ } R_2 \rangle$, where the given presentations are standard. The private key is a homomorphism $\alpha$ of groups that sends generators in $S_1$ to generators in $S_2$, and (commutativity) relations from $R_1$ to relations in $R_2$.

\item Alice selects another graph group $G$ with standard presentation $G=\langle S \textrm{ } | \textrm{ } R \rangle$ and a homomorphism $\beta:G\rightarrow G_1$ sending $S$ to $S_1$ and $R$ to $R_1$. The group $G$ is sent to Bob, and $\beta$ is kept secret by Alice.

\item Now Bob picks a random bit $c$ and sends it to Alice. If $c=0$, Alice sends $\beta$ to Bob, who checks if it takes $S$ to $S_1$ and $R$ to $R_1$. In turn, if $c=1$, Alice sends the composite $\alpha\circ\beta:G\rightarrow G_2$ to Bob, who performs the analogous verification.

\end{itemize}

Observe that, as explained above, the security of this scheme relies in the fact that the Graph Homomorphism problem is NP-complete if the graph in the right has more than two vertices. It is enough to select the graph groups in the scheme with a sufficient number of generators.

\textbf{Zero-knowledge proofs}. 
Motivated by the paper by Goldreich, Micali, and Wigderson in 1991, proofs that yield nothing but their validity, or all languages in NP have zero-knowledge proof systems, we present a ZKP scheme based on NP-completeness of graph group Hamiltonicity in the sense of \cite{FKK-H}.

As commented above in the Hamiltonicity section, in Flores, Kahrobaei and Koberda \cite{FKK-H} prove that Hamiltonicity in graphs is equivalent to Hamiltonicity in the cohomology algebra over the associated right-angled group. Using this result, the authors formulate a zero-knowledge proof protocol based on linear algebra, which we define now in a sketchy way. More details can be found in that paper.

The protocol starts with a finite graph $\Gamma$ that has exactly one Hamiltonian cycle which is supposed to be very difficult to compute, for example when the graph is large and then a greedy algorithm can be very inefficient. The public data is the triple given by $V=H^1(A(\Gamma),\mathbb{F}_2)$, $W=H^2(A(\Gamma),\mathbb{F}_2)$ and the cup product $q=\cup$, which is a Hamiltonian vector space. Note that the coefficients are taken in the field of two elements, in order to make the computations easier. We assume that the generators of the cohomology are given in terms of duals of standard generators of the group (for $H^1$) and their cup products (for $H^2$).

Alice is supposed to have a list $\{v_1^*,\ldots,v_n^*\}$ of standard basis vectors for $V$ such that $q(v_i^*,v_{i+1}^*)\neq 0$ for all $i$ and $q(v_n^*,v_1^*)\neq 0$, and a subset $Y\subset \mathrm{GL}_n(\mathbb{F}_2)$ of
reasonable size (say polynomial in $n$). Moreover, for each $A\in Y$, she knows a Hamiltonian cycle in the complement of the 2-row graph $\mathcal{G}^c(A)$ (see definition in \cite{FKK-H}). The set $Y$ may be public. In turn, Bob may generate unbiased random bits. Now we can define the protocol.
\begin{itemize}
\item
Alice chooses in a random way an element $A\in Y$, obtaining a new basis $\{x_1,\ldots,x_n\}$ from $\{v_1^*,\ldots,v_n^*\}$ using $A$.
Now the knowledge of Hamiltonian cycles in $\Gamma$ and in $\mathcal{G}^c(A)$ makes her
able to find in an efficient way a permutation $\sigma\in S_n$ such that $q(x_{\sigma(i)},x_{\sigma(i+1)})\neq 0$ for $1\leq i\leq n$,
where the indices are
considered cyclically. Alice then creates locked boxes
$\{B_i\}_{1\leq i\leq n}$ for the basis vectors. For each
 pair $\{i,j\}$ with $i<j$, she creates two boxes $N_{i,j}$ and
$S_{i,j}$, where she respectively records the pairing $q(x_i,x_j)\in W$, and $1$ if the entry in $N_{i,j}$ is non-zero and
and $0$ otherwise. In another box $T$, she hides the linear map $A$.
\item
Now Bob takes a random bit and shares it with Alice. If it is $1$, then Alice unlocks the boxes $\{B_i\}_{1\leq i\leq n}$ and the boxes
$\{S_{\sigma(i),\sigma(i+1)}\}_{1\leq i\leq n}$, where again the indices are considered cyclically. Now Bob checks that Alice has produced a cycle in this way.
On the other hand, if the bit is $0$ then Alice opens \[\{B_i\}_{1\leq i\leq n}, \quad \{N_{i,j}\}_{1\leq i<j\leq n},\quad T,\] and Bob recovers the triple $(V,W,q)$.
\end{itemize}

Observe that this protocol may be repeated multiple times, and that it succeeds if Alice correctly complies with all of Bob's requests, and does not succeed if she fails to comply at least once. It can be seen in turn that the protocol is zero-knowledge, and a simulator can be constructed in the same way as Blum in 1987 does.

\textbf{Prospective work}. As said above, the definition of a graph group out of a graph provides an interesting correspondence between algorithmic problems for graphs and groups. In particular, different well-known problems in Graph Theory admit natural counterparts in groups that have not been investigated so far. They may provide in the future new crypto applications, else by using the graph group and a standard presentation as data, and/or defining the group property that models the corresponding math property. Due to limitations of space we only offer here a small list that such problems, more information can be found in \cite{FKK}.
These problems include the vertex cover problem, the clique problem, the independent set problem, the snake-in-the-box problem, the arboricity problem, and the subdivision problem.

From a different point of view, it is worth mentioning work of Chatterji, Kahrobaei et al in 2017, who define different versions of two cryptographic protocols out of the existence of a distorted subgroup $H<G$ inside of a finitely generated group. In the construction of the first of these protocols it is required that the Geodesic Length problem is solvable for $H$ and $G$ in polynomial time, and the Membership problem is also solvable for $H$. In that paper hyperbolic and free-by-cyclic groups are proposed as platforms for the protocol, while in subsequent work of Kahrobaei and Mallahi-Karai in 2019 arithmetic groups are proposed. Following work of Flores, Kahrobaei and Koberda \cite{FKK}, it is possible to construct distorted subgroups inside graph groups such that the Geodesic Length Problem is solvable for them in polynomial time. Moreover, as said above, these authors prove that the Membership problem is solvable for this group in exponential time, and they conjecture that it is likely that the complexity is in fact polynomial. If this happened, then graph groups would become a good platform for this protocol.

\section{Combinatorial Algebra}
There are other combinatorial algebraic problems used for cryptography. Among them, we focus particularly, on one fully homomorphic encryption (FHE) scheme proposal which has been patented \cite{FH} and is currently being used for real life applications, including data analysis over encrypted medical and bioinformatics data \cite{WNK}.
Broadly, homomorphic encryption enables computation over encrypted data. A scheme
is called additively (or multiplicatively) homomorphic if
an encryption scheme is additively homomorphic, then
encryption followed by homomorphic addition is equal to addition followed by encryption.

\subsection{Homomorphic Machine Learning}

Machine learning and statistical techniques are powerful tools for analyzing large amounts of medical and genomic data. On the other hand, ethical concerns and privacy regulations prevent free sharing of this data. Encryption techniques such as fully homomorphic encryption (FHE) enable evaluation over encrypted data. Using FHE, machine learning models such as deep learning, decision trees, and naive Bayes have been
implemented for privacy-preserving applications using medical data. These applications include classifying
encrypted data and training models on encrypted data. FHE has also been shown to enable secure genomic
algorithms, such as paternity and ancestry testing and privacy-preserving applications of genome-wide
association studies, \cite{WNK}

Homomorphic encryption is a form of encryption which allows various
types of computations to be carried out on ciphertext and generate
an encrypted result which, when decrypted, matches the result of
operations performed on the plaintext. Homomorphic encryption  allows, in particular,  chaining together different services without exposing the data to each of those services; this property is important to blockchain technology. 

There are several known cryptosystems (e.g. unpadded RSA, ElGamal,
Goldwasser-Micali) that  allow homomorphic computation of only one
operation (either addition or multiplication) on plaintexts.  A
cryptosystem that supports both addition and multiplication (thereby
preserving the ring structure of the plaintexts) is known as fully
homomorphic encryption (FHE) and is far more powerful. Using such a
scheme, any circuit can be homomorphically evaluated, effectively
allowing the construction of programs which may be run on
encryptions of their inputs to produce an encryption of their
output.  
A fully homomorphic encryption function $E$ encrypts elements of a
ring and respects both ring operations: $E(g_1g_2)=E(g_1)E(g_2)$ and
$E(g_1+g_2)=E(g_1)+E(g_2)$ for any two elements $g_1, g_2$  of the
ring in question. Alternatively, one can encrypt boolean circuits,
and then a fully homomorphic encryption function $E$ should respect
both AND  and OR operations.
The most widely known existing solution to the homomorphic
encryption problem appeared originally in the thesis of Craig Gentry,
was subsequently improved, and the relevant software is
currently being developed by IBM. The security of this solution relies on variants of the ``bounded-distance decoding"
problem that has the property of ``random self-reducibility", which
basically means that it is about as hard on average as it is in the
worst case. While this property is indeed a good evidence of
security, the resulting homomorphic encryption algorithm is too
inefficient to be practical. Very informally, the reason is that, to
provide semantic security, encryption has to be randomized, but on
the other hand, a homomorphism  should   map zero to zero. To
resolve this conflict, the ciphertext zero is ``masked by noise".
The problem now is that during any computation on encrypted data,
this ``noise" tends to accumulate and has to be occasionally reduced
by recryption (also known as {\it bootstrapping}), a process that
produces the equivalent ciphertext but with less noise. Recryption
is an expensive procedure, and its results in real-life computation with this method (or a similar one) are prohibitively slow.
\subsection{An efficient and secure FHE scheme}
Kahrobaei, Shpilrain, Grigov and Lam \cite{FH}, proposed an efficient FHE scheme using combinatorial algebra. Here we give some ideas about the scheme.

We emphasize that this FHE is private-key. 
\begin{enumerate}

\item Plaintexts are elements of a (private) ring $R$.

\item Ciphertexts are elements of a public ring $S$, such that $R
\subset S$ is a subring of $S$. The ring $S$ also has a (private)
ideal $I$ such that $S/I=R'$, where the ring $R'$ is isomorphic to
$R$. (The ring $R'$ may be just equal to $R$, in which case $R$ is
called a {\it retract} of  $S$.)

\item Given $u \in R$,the encryption is $E(u) = u + E(0)$, where $E(0)$ is
a random element of the private ideal $I$ of the ring $S$.

This encryption function is a homomorphism; it obviously respects
addition, and for multiplication we have: $E(u)E(v) = (u + j_1)(v +
j_2) = uv + j_1u + uj_2 + j_1j_2 = uv + j_3 = E(uv)$, where $j_1,
j_2, j_3 \in I$.

\item Private decryption key is a map $\rho$ from $S$ to $R'$  that
takes every element of $I$ to 0, followed by an isomorphism
$\varphi: R' \to R$.
\end{enumerate}

Here is a diagram to ``visualize" this general scheme:

$$R \xrightarrow{E} S \xrightarrow{\rho}
R'  \xrightarrow{\varphi} R.$$

\noindent Note that when we say ``a public ring $S$", this means
that we give to the public a collection of rules for adding and
multiplying elements of $S$. Typically, this can be a (finite) set
of elements that span $S$ as a linear vector space over some $\mathbb{Z}_N$,
together with the multiplication table for $S$ with respect to this
set of elements.

Below is a diagram of the whole encryption process starting with a
real-life database $D$,

$$D \xrightarrow{\alpha}   R \xrightarrow{E} S \xrightarrow{\rho}
R'  \xrightarrow{\varphi} R \xrightarrow{\beta} D,$$

\noindent  where $\beta(\alpha(x)) =x$ for any $x \in D$.
\section{Open problems}
To finish our exposition and at the same time motivate the interested reader, we review several important algorithmic group-theoretic problems motivated by cryptography:
\begin{itemize}
    \item Solving the Hidden Subgroup Problem for various classes of groups. Different instances of groups, mainly finite, have already been considered in this context, namely abelian groups, dihedral groups, symmetric groups, wreath products or the Heisenberg group.
    \smallskip
    \item Complexity analysis of various algorithmic group theoretic problems used in cryptography. According to above, both efficiency and non-efficiency results can be useful in the context, as depending on the situation we may be interested in quick or very difficult decryption.
    \smallskip
    \item Designing machine learning algorithms to solve the algorithmic problems in group theory. This gives rise to heuristic algorithms for the cryptanalysis.
    \smallskip
    \item Cryptographic security analysis for the proposed group-based cryptosystems, including study, simulation and prevention of the possible attacks that the cryptosystem can suffer.
    \smallskip
    \item Searching for more group-based cryptosystems.
    \smallskip
    \item Implementation of the proposed group-theoretic cryptosystems for the real life applications.
\end{itemize}


\textbf{Acknowledgements} We note that because of the limitation of including citations in this forum we have omitted many references by the colleagues in this area. We note such references have been included in the references we mentioned here. This research has been supported in various forms by grants from NSF, ONR, New Frontiers in Research Fund–Exploration(Canada), AAAS, LMS over the years by DK. DK is grateful to the Institut des Hautes \'Etudes Scientifiques for providing excellent scientific environment to do mathematics, this article was finished during DK's visit to IHES.
DK thanks her long term collaborator and mentor, Vladimir Shpilrain for the fruitful collaborations over the last years in direction of group-based cryptography. We thank Christopher Battarbee, Thomas Koberda and Vladimir Shpilrain, who kindly read this manuscript carefully and offered helpful comments. DK's photo is credited to the artist Carlo Cozzolino. The painting is credited to Ms. Sima Hayati (DK's mother).

\bibliographystyle{plain}
\bibliography{bib}

\noindent
\begin{figure}[h!]
    \centering
     \includegraphics[scale=0.3]{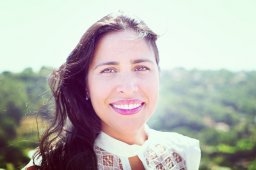}
 \includegraphics[scale=0.2]{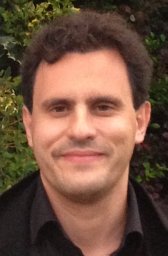} 
 \includegraphics[scale=0.09]{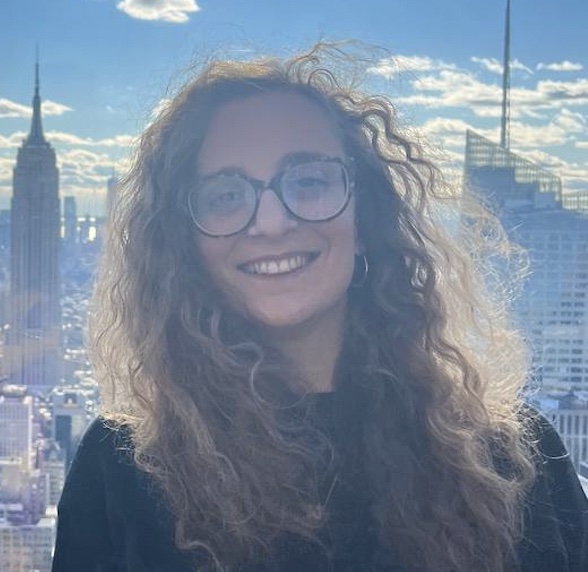}
 \caption{D. Kahrobaei, R. Flores, M. Noce}
\end{figure}

 
\end{document}